\documentclass{PoS}
\usepackage{tabularx,pstricks,array,graphics}
\usepackage{graphicx,epsfig,amsmath,amssymb}
\usepackage{citesort,axodraw,feynmf}
\usepackage{times,mathptmx,fontenc,color}

\def\Journal#1#2#3#4{{#1} {\bf #2}, #3 (#4)}

\def\NPB{{\em Nucl. Phys.} B}
\def\PLB{{\em Phys. Lett.}  B}
\def\PRL{\em Phys. Rev. Lett.}
\def\PRD{{\em Phys. Rev.} D}

\def\be{\begin{equation}}
\def\ee{\end{equation}}
\def\bea{\begin{eqnarray}}
\def\eea{\end{eqnarray}}

\def\lgl{\langle}
\def\rgl{\rangle}

\PoS{PoS(LAT2005)218}
\title{$B_s-\overline{{B}_s}$ mixing with a chiral light quark action }

\ShortTitle{$B_s-\overline{B_s}$ mixing with a chiral light quark action }

\author{D.~Be\'cirevi\'c, \speaker{B.~Blossier}, Ph.~Boucaud, 
A.~Le Yaouanc, J.P.~Leroy and O.~ P\`ene\\
Laboratoire de Physique Th\'eorique, Universit\'e Paris-Sud, 
91405 Orsay Cedex, France\\
E-mail: \email{Benoit.Blossier@th.u-psud.fr}}

\abstract{We study the $B^0_s-\overline{B^0_s}$ mixing amplitude in Standard Model by 
computing  the relevant hadronic matrix element in the 
static limit of lattice HQET with the Neuberger light quark action. In the quenched 
approximation, and after matching to the $\overline{\rm MS}$ scheme in QCD, we obtain 
$B^{\overline{\rm MS}}_{B_s}(m_b)=0.940(16)(22)$.}

\FullConference{XXIIIrd International Symposium on Lattice Field Theory\\

		 25-30 July 2005\\

		 Trinity College, Dublin, Ireland}

\begin{document}
\section{Introduction}
\noindent $B^0_s-\overline{B^0_s}$ mixing is highly important in testing the Standard 
Model (SM) and constrains strongly its extensions. Since it is a flavor changing 
neutral process it occurs through loops so that the corresponding mixing amplitude 
is a sensitive
 measure of $|V_{ts}|$ and $|V_{tb}|$, as the major SM loop contribution comes from 
 $t$-quark. The mixing of weak interaction eigenstates $B^0_s$ and 
$\overline{B^0_s}$ induces a mass gap $\Delta M_s$ between the mass eigenstates 
$B_{sH}$ and $B_{sL}$. Experimentally, only a lower bound to $\Delta M_s$ is currently known, 
namely $\Delta M_s > 14.4 \; \rm{ps}^{-1}$ \cite{PDG}, and the hope is that experimenters 
will soon provide us with an accurate measurement.  

\noindent Theoretically the $B^0_s-\overline{B^0_s}$ mixing is described by means of an Operator 
Product Expansion, $i.e.$ the Standard Model Lagrangian ${\cal L}_{SM}$ is reduced to an effective 
Hamiltonian ${\cal H}_{eff}^{\Delta B=2}$, up to negligible terms of ${\cal O}(1/M^2_W)$:
\be
{\cal H}^{\Delta B=2}_{eff}=\frac{G^2_F}{16\pi^2}M^2_W 
(V^*_{tb}V_{ts})^2\eta_B S_0(x_t) C(\mu_b)Q^{\Delta B=2}_{LL}(\mu_b), \quad 
Q^{\Delta B=2}_{LL}=\bar{b}\gamma_{\mu L}s\; \bar{b}\gamma_{\mu L}s, \quad 
\mu_b\sim m_b
\ee
where $\eta_B=0.55\pm 0.01$ $S_0(x_t)$ is a known Inami-Lim function of $x_t=m^2_t/M^2_W$
\cite{inami}, 
$C(\mu_b)$ is the Wilson coefficient computed perturbatively to NLO in $\alpha_s(\mu_b)$ in the 
$\overline{\rm MS}$ (NDR) scheme, and $Q^{\Delta B=2}_{LL}$ is a 
four-fermions operator coming from the reduction of the box diagrams in 
${\cal L}_{SM}$ to a local operator in the effective theory. 
The hadronic matrix element of $Q^{\Delta B=2}_{LL}$ is conventionally parameterized as
\be
\lgl\overline{B^0_s}|Q^{\Delta B=2}_{LL}(\mu_b)|B^0_s\rgl \equiv \frac 8 3 m^2_{B_s} 
f^2_{B_s}B_{B_s}(\mu_b)\; ,
\ee
where $B_{B_s}(\mu_b)$ is the $B_s$ meson bag parameter and $f_{B_s}$ its decay constant.

\noindent So far $B_{B_s}(\mu_b)$ has been computed by using lattice QCD \cite{bernard}-\cite{aoki2}. 
One of the major problems with those computations is in the following: the standard Wilson light quark 
lattice action breaks explicitely the chiral symmetry, 
which tremendously complicates the renormalization procedure of $Q^{\Delta B =2}_{LL}$ and its matching 
to the continuum. To get around that problem we compute $B_{B_s}(\mu_b)$ by using the lattice 
formulation of QCD in which the chiral symmetry is preserved at finite lattice spacing \cite{overlap}. 
On the other hand, it should be stressed that our heavy quark is static, as 
the currently available lattices do not allow to work directly with the propagating $b$ quark. Thus
our results will suffer from $1/m_b$-corrections.

\section{Computation on the lattice}

\noindent In our numerical simulation we choose to work with the action $S=S^{\rm EH}_h+S^{\rm N}_l$,
where 
\be\nonumber
S^{\rm EH}_h=a^3\sum_{x} 
\left\{\bar{h}^+(x)\left[h^+(x)-V^{{\rm HYP}\dag}_0(x-\hat{0})h^+(x-\hat{0})\right]-
\bar{h}^-(x)\left[V^{{\rm HYP}}_0(x)h^-(x+\hat{0})-h^-(x)\right]\right\}
\ee 
is the static limit of HQET action \cite{eichten} which has been modified after using the so-called 
HYP (hypercubic blocking) procedure \cite{hyp}, that is enough to substantially improve the 
signal/noise ratio \cite{Dellamorte} [the field $h^+(h^-)$ annihilates the static heavy quark
(antiquark)]. 
$S^{\rm N}_l=a^3 \sum_x \bar{\psi}(x) D^{(m_0)}_N\psi(x)$ is the overlap light quark action with
\vspace{-0.2cm}
\be\nonumber
D^{(m_0)}_N=\left(1-\frac{1}{2\rho} am_0 \right)D_N+m_0, \quad 
D_N=\frac{\rho}{a}\left(1+\frac{X}{\sqrt{X^\dag X}}\right), \quad X=D_W -\frac{\rho}{a},
\ee
where $D_W$ is the standard Wilson-Dirac operator. The overlap Dirac operator $D^{(m_0)}_N$
verifies the Ginsparg-Wilson relation 
$\{\gamma^5,D^{(m_0)}_N\}=\frac{a}{\rho}D^{(m_0)}_N\gamma^5D^{(m_0)}_N$ and 
the overlap action is invariant under the chiral light quark transformation \cite{luscher}
\vspace{-0.4cm}
\be\nonumber
\psi \to \psi+i\epsilon \gamma^5\left(1-\frac{a}{\rho}D^{(m_0)}_N\right)\psi, \quad 
\bar{\psi}\to \bar{\psi}(1+i\epsilon \gamma^5),
\ee
which is essential to prevent mixing of four-fermion operators of different chirality 
\cite{becirevic}. In other words, in the renormalization 
procedure, the subtraction of the spurious mixing with $d=6$ operators will not be needed.


\begin{figure}
\begin{center}
\vspace{-0.5cm}
\includegraphics*[width=6.5cm, height=4.7cm]{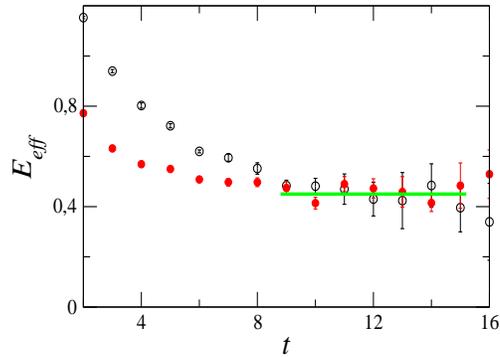}
\end{center}
\vspace{-0.5cm}
\caption{Effective binding energy of the $0^-$-state when currents are local (unfilled symbols) 
or smeared (filled symbols)}
\label{smearing}
\end{figure}
\noindent We thus compute the two- and three-point functions:
\vspace{-0.3cm}
\be\label{c2}
\tilde{C}^{(2)\pm}_{AA}(t)=\lgl\sum_{\vec{x}}\tilde{A}^\pm_0(\vec{x},t)\tilde{A}^{\pm\dag}_0(0)\rgl_{_{U}}
\stackrel{t\gg 0}{\longrightarrow} \tilde{Z}_A e^{-\epsilon t},\\
\ee
\vspace{-0.7cm}
\bea\label{c3}
\tilde{C}^{(3)}_{VV+AA}(t_i,t)=\lgl\sum_{\vec{x},\vec{y}} 
\tilde{A}^+_0(\vec{x},t_i)\tilde{O}_1(0,0)\tilde{A}^{-\dag}_0(\vec{y},t)\rgl_{_{U}}
&\stackrel{t_i-t \gg 0}{\longrightarrow}&\tilde{Z}_A\, _v\lgl
\overline{B_s}|\tilde{O}_1(\mu)|B_s\rgl_v\; e^{-\epsilon(t_i-t)},\\
\label{c32}
\tilde{C}^{(3)}_{SS+PP}(t_i,t)=\lgl\sum_{\vec{x},\vec{y}} 
\tilde{A}^+_0(\vec{x},t_i)\tilde{O}_2(0,0)\tilde{A}^{-\dag}_0(\vec{y},t)\rgl_{_{U}}
&\stackrel{t_i-t \gg 0}{\longrightarrow}& \tilde{Z}_A\,\, _v\lgl \overline{B_s}|
\tilde{O}_2(\mu)|B_s\rgl_v\; e^{-\epsilon(t_i-t)},
\eea
\vspace{-0.4cm}
\be\nonumber
\tilde{A}^\pm_0\equiv \bar{h}^\pm\gamma_0\gamma^5 s, \quad
\tilde{O}_1=\bar{h}^{(+)i}\gamma_\mu(1-\gamma^5)s^i\bar{h}^{(-)j}\gamma_\mu(1-\gamma^5)s^j, \quad
\tilde{O}_2=\bar{h}^{(+)i}(1-\gamma^5)s^i\bar{h}^{(-)j}(1-\gamma^5)s^j.
\ee
$\sqrt{\tilde{Z}_A}=\lgl 0|\tilde{A}^-_0|B_s\rgl_v
=\lgl 0|\tilde{A}^+_0|\overline{B_s}\rgl_v$ and $\epsilon$ is the binding energy of the pseudoscalar 
heavy-light meson. In $\tilde{C}^{(2)\pm}(t_i,t)$ one current $\tilde{A}^\pm_0$ is local whereas 
the other is smeared. 
The role of the smearing is to isolate earlier the ground state \cite{boyle}, as shown in Fig.
\ref{smearing}~\footnote{Even if the time interval from which we extract the binding energy 
starts at $t=9$ (green line), the overlap with radial excitations is
quite reduced since $t=6$ when currents are smeared.}. We see that the same state is isolated when 
purely local currents are used (with those currents the signal does not exist if 
$V^{\rm HYP}_0$ is not used in the heavy quark action).
The source operators in $\tilde{C}^{(3)}_{VV+AA}(t_i,t)$ and $\tilde{C}^{(3)}_{SS+PP}(t_i,t)$ 
are the smeared currents $\tilde{A}^\pm_0$, whereas the four-fermion operators $\tilde{O}_1$ and 
$\tilde{O}_2$ are purely local.
In (\ref{c2}), (\ref{c3}) and (\ref{c32}) the subscript "$v$" and superscript "$\sim$" are designed 
to remind the reader that states and operators are defined in HQET. Note that in the 
computation of $\tilde{C}^{(3)}_{VV+AA}(t_i,t)$ and $\tilde{C}^{(3)}_{SS+PP}(t_i,t)$ there
are two terms, coming from 
two different Wick contractions, namely $\displaystyle{\sum_i} B_{ii}(t)\displaystyle{\sum_j}B_{jj}
(t_i)$ and $\displaystyle{\sum_{i,j}}B_{ij}(t)B_{ji}(t_i)$, where $i$, $j$ are the color indices and
$B_{ij}(t)={\rm Tr}\left[\sum_{\vec{x}} \gamma_{\mu L}{\cal S}^{\dag ik}_L(0;\vec{x},t) \gamma_0\gamma^5
{\cal S}^{kj}_H(\vec{x},t;0)\right]$; ${\cal S}_L$ and ${\cal S}_H$ are the light and heavy propagators 
respectively and the trace is over spinor indices.

\begin{figure}
\begin{center}
\vspace{-0.5cm}
\includegraphics*[width=6.5cm, height=4.7cm]{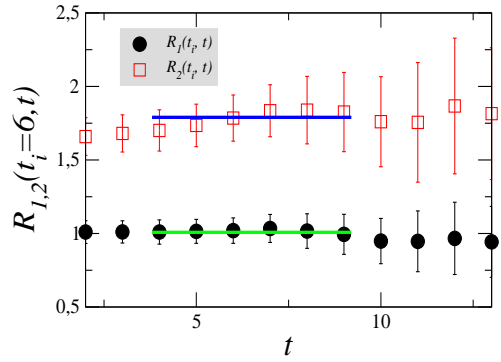}
\end{center}
\vspace{-0.5cm}
\caption{Signals for $R_{1,2}(t_i,t)$ defined in eq $(2.4)$: green and blue lines
indicate the time interval on which we fit the signal to a constant to extract
$\tilde{B}_1(a)$ and $\tilde{B}_2(a)$ respectively}
\label{rap}
\end{figure}

\noindent After having computed the correlation functions (\ref{c2}), (\ref{c3}) and
(\ref{c32}) we build the following two ratios $R_1(t_i,t)$ and $R_2(t_i,t)$:
\bea\label{ratio}
\nonumber
R_1(t_i,t)=\frac{\tilde{C}^{(3)}_{VV+AA}(t_i,t)}{\frac 8 3 \tilde{Z}^2_A \tilde{C}^{(2)+}_{AA}(t_i)
\tilde{C}^{(2)-}_{AA}(t)}\stackrel{t_i-t \gg 0}{\longrightarrow}
\frac{_v\lgl \overline{B_s}|\tilde{O}_1|B_s\rgl_v}{\frac 8 3 |\lgl 0|\tilde{A}^-_0|B_s\rgl_v|^2}
\equiv \tilde{B}_1(a),\\
R_2(t_i,t)=\frac{C^{(3)}_{SS+PP}(t_i,t)}{-\frac 5 3 \tilde{Z}^2_A \tilde{C}^{(2)+}_{AA}(t_i)
\tilde{C}^{(2)-}_{AA}(t)}\stackrel{t_i-t \gg 0}{\longrightarrow}
\frac{_v\lgl \overline{B_s}|\tilde{O}_2|B_s\rgl_v}{-\frac 5 3 |\lgl 0|\tilde{A}^-_0|B_s\rgl_v|^2}
\equiv \tilde{B}_2(a).
\eea
Those ratios are calculated either with a fixed time $t \in [-6,-8,-10,-12,-14,-16]$ and  
$t_i$ free, or by fixing $t_i \in [6,8,10,12,14,16]$ while letting $t$ free. We take the 
average of the two options. In Fig. \ref{rap} we show the quality of the signals for
$R_{1,2}(t_i,t)$, with $t_i=6$ fixed. The signal for $\tilde{B}_1(a)$ is quite
stable as a function of $t_i$, whereas the signal for $\tilde{B}_2(a)$ rapidly deteriorates for 
larger $t_i$, and is completely lost for $t_i>10$.

\section{Extraction of physical $B_{B_s}$}

\noindent Three steps are required to extract $B_{B_s}\equiv B_1$ from the lattice:\\
(1) $\tilde{B}_{1,2}(a)$ are matched onto the continuum $\overline{\rm MS}$(NDR) scheme at NLO in
perturbation theory at the renormalization scale $\mu=1/a$ \cite{becirevic},\\
(2) $\tilde{B}_{1,2}$ are evolved from $\mu=1/a$ to $\mu=m_b$ by using the HQET anomalous
dimension matrix, known to 2-loop accuracy in perturbation theory \cite{spqcdr,run},\\
(3) $\tilde{B}_{1,2}(\mu=m_b)$ are then matched onto their QCD counterpart, $B_{1,2}(m_b)$, 
in the $\overline{\rm MS}$(NDR) scheme at NLO \cite{run}.\\
The advantage of using a chiral light quark action for the step (1) lies in the fact that 
four-fermion operators can mix only with 
a four-fermion operator of the same chirality. In other words we have not more than 4 independent 
renormalization 
constants in the renormalization matrix, because $\tilde{O}_1$ and $\tilde{O}_2$ can mix neither 
with $\tilde{O}_3\equiv \bar{h}^+\gamma_{\mu L}s\;\bar{h}^-\gamma_{\mu R}s$, nor with 
$\tilde{O}_4\equiv \bar{h}^+(1-\gamma^5)s\;\bar{h}^-(1+\gamma^5)s$ :
\be\nonumber
\left(\begin{array}{c}
\tilde{B}^{\overline{\rm MS}}_1(\mu)\\
\tilde{B}^{\overline{\rm MS}}_2(\mu)
\end{array}
\right)
=
\left(\begin{array}{cc}
Z_{11}(a\mu)&Z_{12}(a\mu)\\
Z_{21}(a\mu)&Z_{22}(a\mu)
\end{array}\right)
\left(\begin{array}{c}
\tilde{B}_1(a)\\
\tilde{B}_2(a)
\end{array}\right).
\ee
Actually, because of the heavy quark symmetry, those constants are not all independent. 
By using the heavy quark symmetry (HQS) transformations 
$\bar{h}^{(\pm)}(x)\stackrel{HQS(i)}{\longrightarrow} 
-\frac 1 2 \epsilon^{ijk}\bar{h}^{(\pm)}(x)\gamma_j\gamma_k 
\;\; (i=1,2,3)$, and the equations of motion for the heavy quark 
$\bar{h}^{(\pm)}\gamma_0=\pm \bar{h}^{(\pm)}$, we see that: 
\be\nonumber
O_{SS+PP}\equiv-O_{(VV+AA)_0}, \quad
O_{VV+AA}\stackrel{HQS(i)}{\longrightarrow}O_{VV+AA}, \quad 
O_{SS+PP}\stackrel{HQS(i)}{\longrightarrow}-O_{(VV+AA)_i}.
\ee
As the action is invariant under the HQS transformations, we can deduce important constraints on the
renormalization matrix $Z_{ij}$:
\vspace{-0.2cm}
\bea\nonumber
\lgl O_{VV+AA}(\mu) \rgl&=& Z_{11} \lgl O_{VV+AA}(a) \rgl +Z_{12} \lgl O_{SS+PP}(a) \rgl,\\
\nonumber
\lgl O_{VV+AA}(\mu) \rgl&=& Z_{11} \lgl O_{VV+AA}(a) \rgl -Z_{12} \lgl O_{(VV+AA)_i}(a) \rgl \quad ({\rm
HQS(i)}),
\eea
which implies that $Z_{12}=0$. Moreover
\vspace{-0.2cm}
\bea\nonumber
\lgl O_{SS+PP}(\mu) \rgl&=& Z_{21} \lgl O_{VV+AA}(a) \rgl +Z_{22} \lgl O_{SS+PP}(a) \rgl,\\
\nonumber
-\lgl O_{(VV+AA)_i}(\mu) \rgl&=& Z_{21} \lgl O_{VV+AA}(a) \rgl -Z_{22} \lgl O_{(VV+AA)_i}(a) 
\rgl \quad ({\rm HQS(i)}),
\eea

\vspace{-0.8cm}
\bea\nonumber
-\sum_{i=1,3} O_{(VV+AA)_i}(\mu) \pm O_{(VV+AA)_0}(\mu) &\equiv& -\lgl O_{SS+PP}(\mu) \rgl -
\lgl O_{VV+AA}(\mu) \rgl,\\ 
\nonumber
&=& (3Z_{21}-Z_{22}) \lgl O_{VV+AA}(a) \rgl  - Z_{22}\lgl O_{SS+PP}(a)\rgl,\\
\nonumber
&=& -(Z_{11}+Z_{21}) \lgl O_{VV+AA}(a) \rgl - Z_{22} \lgl O_{SS+PP}(a)\rgl,
\eea
which leads to $Z_{21}=(Z_{22}-Z_{11})/4$.\\
Finally we have 
\vspace{-0.2cm}
\be\nonumber
\left(\begin{array}{c}
\tilde{B}^{\overline{\rm MS}}_1(\mu)\\
\tilde{B}^{\overline{\rm MS}}_2(\mu)
\end{array}
\right)
=
\left(\begin{array}{cc}
Z_{11}(a\mu)&0\\
\frac 1 4 \left[Z_{22}(a\mu)-Z_{11}(a\mu)\right]&Z_{22}(a\mu)
\end{array}\right)
\left(\begin{array}{c}
\tilde{B}_1(a)\\
\tilde{B}_2(a)
\end{array}\right)
\ee
Therefore only two independent renormalization constants are required to match the bag parameters
$\tilde{B}_{1,2}(a)$ computed on the lattice to their counterpart renormalized in $\overline{\rm MS}$
scheme.

\section{Results and discussion}

\begin{figure}
\begin{center}
\begin{picture}(180,30)(60,-10)

\LinAxis(0,40)(160,40)(3,4,-3.2,0,0.2)
\LinAxis(0,-140)(160,-140)(3,4,3.2,0,0.2)
\Line(0,40)(0,-140) \Line(160,40)(160,-140)
\DashLine(-20,-110)(160,-110){2}

\Text(0,-150)[]{0.7}
\Text(40,-150)[]{0.8}
\Text(80,-150)[]{0.9}
\Text(120,-150)[]{1.0}
\Text(160,-150)[]{1.1}

\Text(20,30)[]{\small{$0.97(12)$}}
\Text(20,10)[]{\small{$0.81(7)$}}
\Text(20,-10)[]{\small{$0.93(^{+08}_{-10})$}}
\Text(20,-30)[]{\small{$0.90(^{+4}_{-2})$}}
\Text(20,-50)[]{\small{0.87(2)}}
\Text(20,-70)[]{\small{0.85(5)}}
\Text(20,-90)[]{\small{$0.85(6)$}}
\Text(20,-120)[]{\small{$0.94(3)$}}


\Line(90,-120)(110,-120)
\Line(90,-122)(90,-118)
\Line(110,-122)(110,-118)
\CTri(97,-120)(100,-116)(100,-124){Blue}{Blue}
\CTri(100,-116)(100,-124)(103,-120){Blue}{Blue}

\Line(38,-100)(81,-100)
\Line(38,-102)(38,-98)
\Line(81,-102)(81,-98)
\CBoxc(60,-100)(4,4){Red}{Red}

\Line(40,-80)(80,-80)
\Line(40,-82)(40,-78)
\Line(80,-82)(80,-78)
\CBoxc(60,-80)(4,4){Red}{Red}

\Line(60,-60)(80,-60)
\Line(60,-62)(60,-58)
\Line(80,-62)(80,-58)
\CTri(67,-63)(70,-56)(73,-63){Red}{Red}


\SetColor{Black}
\Line(72,-40)(95,-40)
\Line(72,-42)(72,-38)
\Line(95,-42)(95,-38)
\CCirc(80,-40){3}{Red}{Red}

\SetColor{Black}
\Line(52,-20)(122,-20)
\Line(52,-22)(52,-18)
\Line(122,-22)(122,-18)
\CCirc(92,-20){3}{Red}{Red}

\SetColor{Black}
\Line(12,0)(72,0)
\Line(12,-2)(12,2)
\Line(72,-2)(72,2)
\CCirc(42,0){3}{Blue}{Blue}

\SetColor{Black}
\Line(70,20)(158,20)
\Line(70,18)(70,22)
\Line(158,18)(158,22)
\CCirc(114,20){3}{Red}{Red}

\Text(168,-120)[l]{Orsay, static heavy quark (2005)}

\Text(168,-100)[l]{JLQCD (unq. $N_f=2$), NRQCD (2003)}

\Text(168,-80)[l]{JLQCD, NRQCD (2003)}

\Text(168,-60)[l]{SPQcdR (2002)}

\Text(168,-40)[l]{UKQCD (2001)}

\Text(168,-20)[l]{APE (2000)}

\Text(168,0)[l]{Gimenez and Reyes (1999)}

\Text(168,20)[l]{Bernard, Blum and Soni (1998)}

\rText(-10,-30)[][l]{\textcolor{blue}{\small{Renormalization with subtractions}}}
\rText(-14,-125)[][l]{\textcolor{blue}{\small{no}}}
\rText(-7,-125)[][l]{\textcolor{blue}{\small{subtr.}}}

\end{picture}
\end{center}
\vspace{4.6cm}
\caption{ \label{comp} Various lattice values of $B^{\overline{MS}}_{B_s}(m_b)$
\cite{bernard}-\cite{aoki2}; blue symbols correspond to a computation made with a static heavy 
quark}
\end{figure}
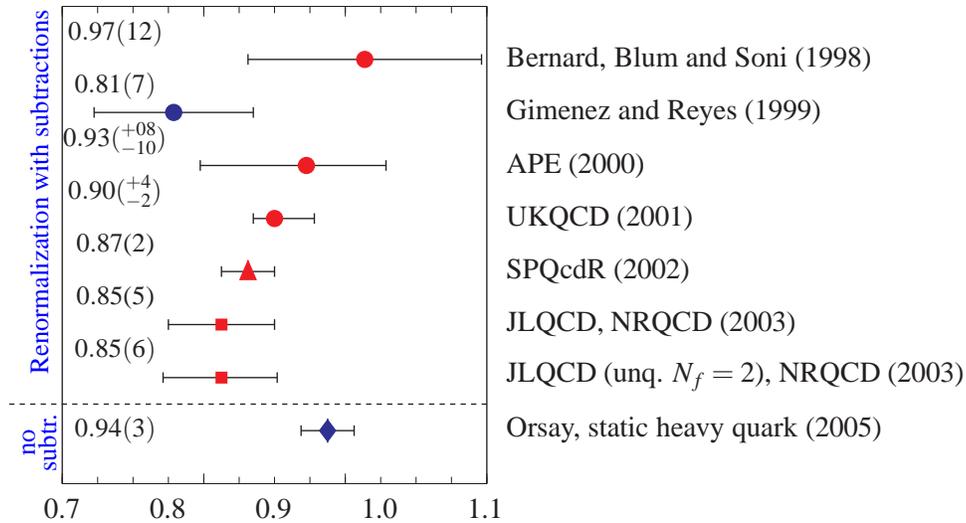

\noindent Our results are based on two simulations, with the parameters given in Tab. \ref{table}.
\begin{table}
\begin{center}
\vspace{-0.5cm}
\begin{tabular}{|c|c|c|c|c|}
\hline
$\beta$&V&${\rm N_{conf}}$&$\rho$&$m^s_0$\\
\hline
6.0&$16^3\times 32$&80&1.4&0.06 GeV\\
\hline
5.85&$16^3\times 32$&30&1.6&0.09 GeV\\
\hline
\end{tabular}
\end{center}
\vspace{-0.5cm}
\caption{Parameters of our simulations: $m^s_0$ and $\rho$ have been chosen following
\cite{giusti,hernandez}}
\label{table}
\end{table} 
We find $B^{\overline{\rm MS}}_{B_s}(m_b)=0.940(16)(22)$, where the first error is statistical,
the second is systematic and contains the error from the estimation of $\alpha_s(1/a)$ and the finite 
$a$ effects. From Fig. \ref{comp} 
it can be seen that our value is larger than the previous static
result \cite{gimenez}. This difference is likely due to the use of 
Neuberger light quark action (no subtractions), due to the use of the HYP procedure, or
the combination of both. From Fig. \ref{comp} we also notice that our value is also somewhat 
larger than the results obtained with the 
propagating heavy quark, which is due to our neglect of $1/m_b$ 
corrections or their not so proper renormalization. JLQCD collaboration showed that the errors due to 
quenching are likely to be small \cite{aoki1,aoki2}. We also plan to address that issue by unquenching 
the $B^0_s-\overline{B^0_s}$ mixing amplitude 
in the static limit and by avoiding the subtraction procedure as well. The feasibility study 
by means of twisted mass QCD is underway.

\end{document}